\documentclass[12pt,preprint]{aastex}

\begin{document}

\title{Photometric study of three ultrashort-period contact binaries}

\author{L. Liu\altaffilmark{1,2,3,4},
S.-B. Qian\altaffilmark{1,2,3,4},
E. Fern\'{a}ndez Laj\'{u}s\altaffilmark{5,6,7},
A. Essam\altaffilmark{8},
M. A. El-Sadek\altaffilmark{8}, and
X. Xiong\altaffilmark{1,2,3,4}}\singlespace

\altaffiltext{1}{Yunnan Observatories, Chinese Academy of Sciences, 396 Yangfangwang, Guandu District, Kunming, 650216, P. R. China (e-mail: LiuL@ynao.ac.cn)}
\altaffiltext{2}{Key Laboratory for the Structure and Evolution of Celestial Objects, Chinese Academy of Sciences, 396 Yangfangwang, Guandu District, Kunming, 650216, P. R. China}
\altaffiltext{3}{Center for Astronomical Mega-Science, Chinese Academy of Sciences, 20A Datun Road, Chaoyang District, Beijing, 100012, P. R. China}
\altaffiltext{4}{University of Chinese Academy of Sciences, Yuquan Road 19\#, Sijingshang Block, 100049 Beijing, China}
\altaffiltext{5}{Facultad de Ciencias Astron\'{o}micas y Geof\'{\i}sicas, Universidad Nacional de La Plata, 1900 La Plata, Buenos Aires, Argentina}
\altaffiltext{6}{Instituto de Astrof\'{\i}sica de La Plata (CCT La Plata-CONICET, UNLP), Argentina}
\altaffiltext{7}{Visiting Astronomer, Complejo Astron\'{o}mico El Leoncito, operated under agreement
between the Consejo Nacional de Investigaciones Cient\'{i}ficas y T\'{e}cnicas de la Rep\'{u}blica
Argentina and the National Universities of La Plata, C\'{o}rdoba, and San Juan}
\altaffiltext{8}{National Research Institute of Astronomy and Geophysics, Department of Astronomy, Helwan, Cairo, Egypt}

\begin{abstract}
We carried out high-precision photometric observations of three eclipsing ultrashort-period contact binaries (USPCBs). Theoretical models were fitted to the light-curves by means of the Wilson-Devinney code. The solutions suggest that the three targets have evolved to a contact phase. The photometric results are as follows:
a) 1SWASP\,J030749.87$-$365201.7, $q=0.439\pm0.003, f=0.0\pm3.6\,\%$;
b) 1SWASP\,J213252.93$-$441822.6, $q=0.560\pm0.003, f=14.2\pm1.9\,\%$;
c) 1SWASP\,J200059.78$+$054408.9, $q=0.436\pm0.008, f=58.4\pm1.8\,\%$.
The light curves show O'Connell effects, which can be modeled by assumed cool spots. The cool spots models are strongly supported by the night-to-night variations in the $I$-band light curves of 1SWASP\,J030749.87$-$365201.7. For a comparative study, we collected the whole set of 28 well-studied USPCBs with P $<$ 0.24 day. Thus, we found that most of them (17 of 28) are in shallow contact (i.e. fill-out factors $f<20\,\%$). Only 4 USPCBs have deep fill-out factors (i.e. $f>50\,\%$). Generally, contact binaries with deep fill-out factors are going to merge, but it is believed that USPCBs have just evolved to a contact phase. Hence, the deep USPCB 1SWASP\,J200059.78$+$054408.9 seems to be a contradiction, making it very interesting. Particularly, 1SWASP\,J030749.87$-$365201.7 is a zero contact binary within thermal equilibrium, implying that it should be a turn-off sample as predicted by the thermal relaxation oscillation (TRO) theory.
\end{abstract}


\section{Introduction}
W UMa-type contact binaries are usually composed by two cool main sequence stars, where both components are covered by a common convective envelope (CCE). According to the period distribution, \citet{Rucinski1992} found a short-period limit (0.22 days) of contact binaries. After then, some papers tried to explain this limitation:
(1) the components become fully convective below a certain period \citep[e.g.][]{Rucinski1992, Rucinski1997, Paczynskietal2006, Beckeretal2011} so that the system becomes unstable;
(2) the timescale of the angular momentum loss (AML) is much longer than the age of the universe, so that the ultrashort-period (period being around 0.22 days) eclipsing binary (USPCB) state cannot be achieved \citep{Stepien2006, Stepien2011}. Almost fully convective structure makes USPCBs to be different from the F, G and K type contact binaries.

In recent years, many USPCBs have just been found with the exoplanet searching projects \citep[a list with the names of these surveys was introducd by][]{Koenetal2016}. \citet{Nortonetal2011} published a catalogue of 53 ultrashort-period eclipsing binary (USPEB) candidates, while \citet{Lohretal2013a} investigated the period changes of 143 USPEBs, based on the data of Super Wide Angle Search for Planets (SuperWASP). \citet{Koenetal2016} confirmed 29 USPEB systems being in overcontact configuration according to a preliminary Fourier decomposition analysis. A series of studies on individual USPCBs have been done in the last 5 years \citep[e.g.][]{DimitrovKjurkchieva2015, Qianetal2015a, Liuetal2015a, Jiangetal2015a}. These studies showed that most USPCBs have shallower fill-out factors ($f<20\,\%$) indicating that USPCBs are in the beginning phase of the contact configuration. Also, some USPCBs have already broken the known short-period limitation or low-mass limitation
\citep[e.g. SDSS\,J001641$-$000925,][]{Davenportetal2013, Qianetal2015b}. Very recently, \citet{Qianetal2017} made a statistics with a sample of 5363 EW-type binary stars which were determined by LAMOST based on good spectroscopic observations, finding a peak of period distribution at 0.29 days. This distribution implies that there should be many short-period contact binaries. Maybe more USPCBs can be found in this sample.

The special USPCBs that are under the critical conditions are interesting and important to understand the evolutionary boundary conditions. Therefore, we observed the following USPCB candidates: 1SWASP\,J030749.87$-$365201.7 (hereafter J030749), 1SWASP\,J213252.93$-$441822.6 (hereafter J213252) and 1SWASP\,J200059.78$+$054408.9 (hereafter J200059), with 2-meter-class telescopes. J030749 was discovered by \citet{Nortonetal2011}, while J213252 and J200059 were discovered by \citet{Lohretal2013a}. The period of these three systems are: 19584.393\,s (0.22667122\,day), 19114.669\,s (0.22123459\,day) and 17771.663\,s (0.20569054\,day), respectively \citep{Lohretal2013a}. In the literature and in the surveys databases, the light curves of these three USPCBs show very large scatters. So, in this paper, we analyzed these systems with our new observed high-precision multi-color light curves.
We report here that the zero contact J030749 and the deep contact J200059, are such special samples mentioned above.

\section{Observation and data reduction}
The southern targets J030749 and J213252 were observed using the 2.15\,m Jorge Sahade telescope (JST) at Complejo Astron\'{o}mico El Leoncito Observatory (CASLEO), San Juan, Argentina. The JST was equipped with a Versarray 2048B-Princeton Instruments CCD camera at the Cassegrain focus, covering a 5$\times$5 arcmin$^2$ field of view. The CCD was cooled with liquid nitrogen and a 5$\times$5 pixel binning factor was applied to enhance the signal-to-noise ratio and to reduce the exposure times. A standard $BVRI$ filter set was used. Thus, direct images of J030749 were acquired from November 24 to December 4, 2014, while new images of J213252 were acquired in September 11, 2015.

The new $VRI$ images of J200059 were taken in August 1, 2016, with the 1.88\,m reflector telescope at the Kottamia Observatory, Astronomy Department, National Research Institute of Astronomy and Geophysics (NRIAG), 11421 Helwan, Cairo, Egypt. A 2048$\times$2048 pixels EEV CCD 42-40 camera was attached at the Newtonian focus of the telescope, reaching a 10$\times$10 arcmin$^2$ field of view.
The CCD was also cooled by liquid nitrogen. A summary of the observations can be found in Table~\ref{tab:summary observation_table}.

Bias subtraction and flat-fielding corrections were applied to the images in the standard way by means of the IRAF
\footnote{IRAF is written and supported by the National Optical Astronomy Observatories (NOAO) in Tucson, Arizona. NOAO is operated by the Association of Universities for Research in Astronomy (AURA), Inc. under cooperative agreement with the National Science Foundation.}
facilities. Subsequent aperture photometry was performed to the targets and the other reference stars to get the differential photometry. The comparison and check stars for the differential photometry are listed in Table~\ref{tab:target comparison_table}. The typical errors in the differential photometry are about 0.009\,mag for J030749, 0.007\,mag for J213252, and 0.006\,mag for J200059.

Finally, the light curves were made up and they are shown in Fig~\ref{fig:light curves_figure}. The phases were calculated with the following ephemeris: a) {\centering $2456994.77031+0^d.22667122{\times}E$} for J030749, b) {\centering $2457276.57884+0^d.22123459{\times}E$} for J213252, and c) {\centering $2457604.37104+0^d.20569054{\times}E$} for J200059.

\section{Light curves solutions}
To determine the photometric elements and to understand the geometrical structure and evolutionary state of these three USPCBs, we analyzed their multi-color light curves using the latest version of the Wilson-Devinney (W-D) code \citep{WilsonDevinney1971, Wilson1979, Wilson1990, Wilson2008, Wilson2012, vanHammeWilson2007, Wilsonetal2010, WilsonvanHamme2014}.

Usually, the effective temperatures are estimated by colors. At the beginning, we estimated the $T_1$ (effective temperature of the hot component) based on the $JHK$ color indexes because our targets have late spectral types (red color). The corresponding $JHK$ magnitudes \citep[2MASS catalogue
\footnote{The online 2MASS catalogue can be found in the website of http://vizier.u-strasbg.fr/viz-bin/VizieR-3?-source=II/246/out\&-out.max=50\&-out.form=HTML\%20Table\&-out.add=\_r\&-out.add=\_RAJ,\_DEJ\&-sort=\_r\&-oc.form=sexa.}
,][]{Cutrietal2003} are listed in Table~\ref{tab:target comparison_table}.

\begin{figure}
\begin{center}
\includegraphics[angle=0,scale=.85]{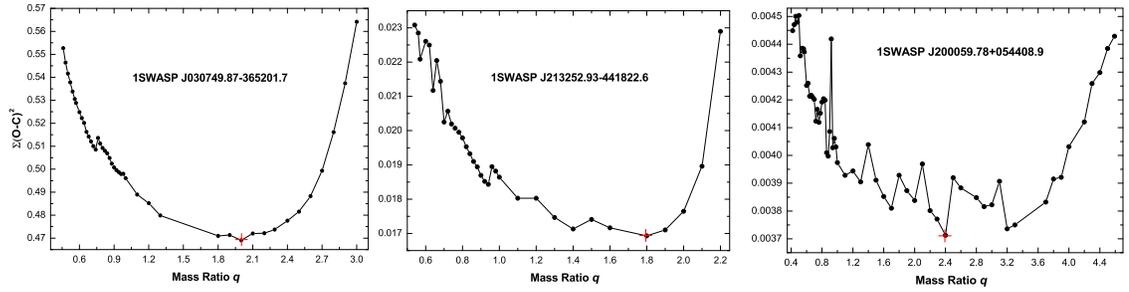}
\caption{The relation between $q$ and fitting residuals for the three USPCBs. The red cross in each diagram presents the initial value of $q$ at the beginning of the W-D program running.}
    \label{fig:q search_figure}
\end{center}
\end{figure}

The effective temperatures are computed by the method of \citet{WortheyLee2011}. Then, we applied the $q$-searching method to find an initial $q$ as the input parameter for the W-D code. Such $q$-searching grid method is just aimed to make the program to converge faster. In other words, the advantage of the $q$-searching method is to save computing time. The results of $q$-searching are displayed in Fig~\ref{fig:q search_figure}. The red crosses are the suggested values of $q$ for each system. For the solutions, the bolometric albedos $A_1=A_2=0.5$ \citep{Rucinski1969} and
gravity-darkening coefficients $g_1=g_2=0.32$ \citep{Lucy1967} were assumed, as it corresponds to a common convective envelope of both components. Square root limb-darkening coefficients were used, according to \citet{ClaretGimenez1990}. The adjustable parameters were: the mass ratio $q$; the orbital inclination $i$; the potential $\Omega$ of both stars; the mean temperature of star\,2, $T_2$; the monochromatic luminosity of star\,1. We adopt mode\,3 in the final solutions, because every solution finally converged to that mode. In the W-D code, mode\,3 is a mode for contact binaries, with the constraint of $\Omega_1=\Omega_2$. The systems which are in geometrical contact without being in thermal contact can be simulated by this mode.

\begin{figure}
\begin{center}
\includegraphics[angle=0,scale=1.5]{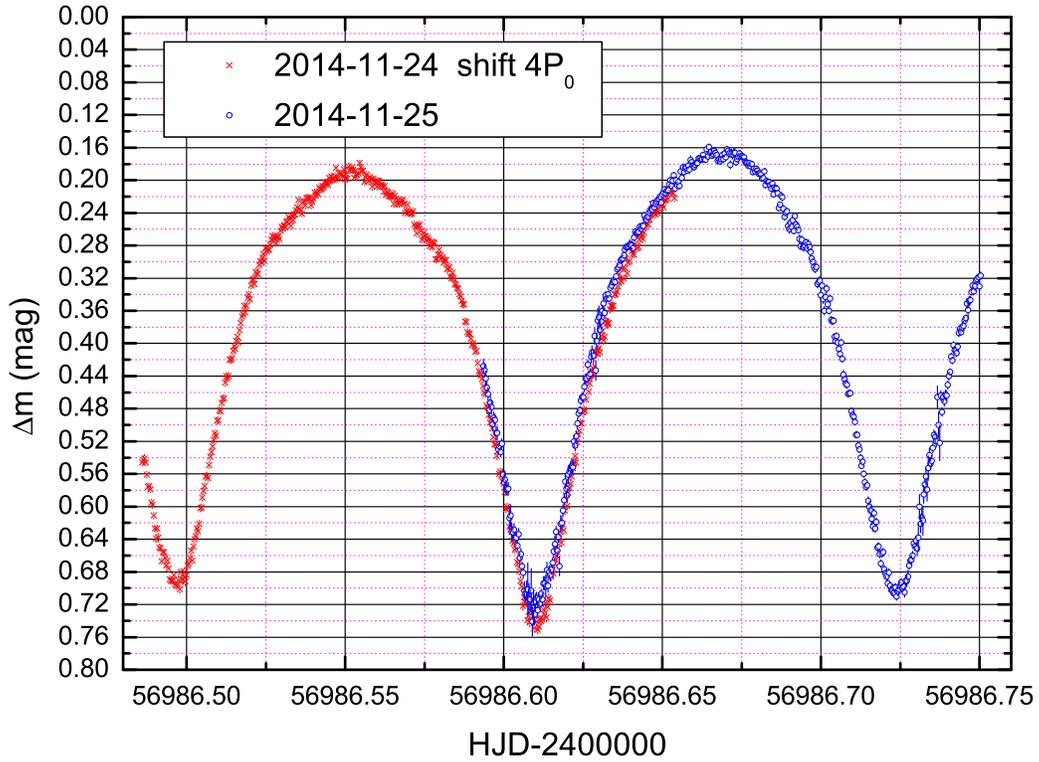}
\caption{The night-to-night light curve variation of 1SWASP\,J030749.87$-$365201.7. The red data points were observed on the night of 2014-11-24 with the I filter, while the blue data points were observed on the night of 2014-11-25 with the same filter. We add four times of $P_0$ (0.2266712 day) to the red points. However, this two parts of light curves do not join well. Moreover, a changing of depths of minima is clearly seen. This phenomenon could be caused by variations of the presented cool spots.}
    \label{fig:night-to-night variation_figure}
\end{center}
\end{figure}

However, all light curves showed O'Connell effects \citep{O'Connell1951}. The presence of cool spots on the more massive component can simulate the light curves very well, and it is supported by:
i) late type stars have deep convective zones so that they have strong magnetic fields; ii) fast rotate late type stars should have stronger magnetic fields than normal \citep{BarnesCollierCameron2001, Barnesetal2004}; iii) the more massive component has a deeper convective zone than the less massive one \citep{Mullan1975}. For J030749 the situations of the cool spots are more complicated, being supported by its $I$-band night-to-night variations in the light curves (Fig~\ref{fig:night-to-night variation_figure}). Three cool spots are added to simulate its light curves. One is on the less massive component and two spots are on the more massive component. So many spots on J030749 suggest that it has very strong magnetic activities. Emphatically, the longitudes of spots can be constrained exactly by the distortions in the light curves. Nevertheless, the other three parameters: latitudes, radii and temperature ratios, are not independent. A very recent study of cool spots on M dwarfs revealed that the fractus cool spots occurred at high latitudes with high frequency \citep{Barnesetal2015}, according to the Doppler maps.

\begin{figure}
\begin{center}
\includegraphics[angle=0,scale=1.3]{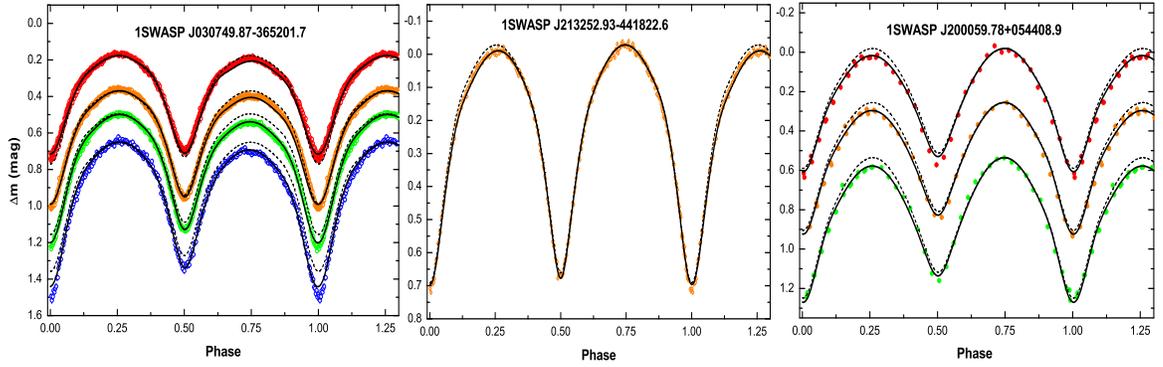}
 \caption{Observed and fitted curves for the three USPCBs. The color points denote the observed data with different filters. Blue denotes B filter; green denotes V filter; orange denotes R filter; red denotes I filter. The black solid lines are the theoretical fittings, with the modeling of dark spots. The dashed lines are the best-fit results without including any spots.}
    \label{fig:light curves_figure}
\end{center}
\end{figure}

\begin{figure}
\begin{center}
\includegraphics[angle=0,scale=.7]{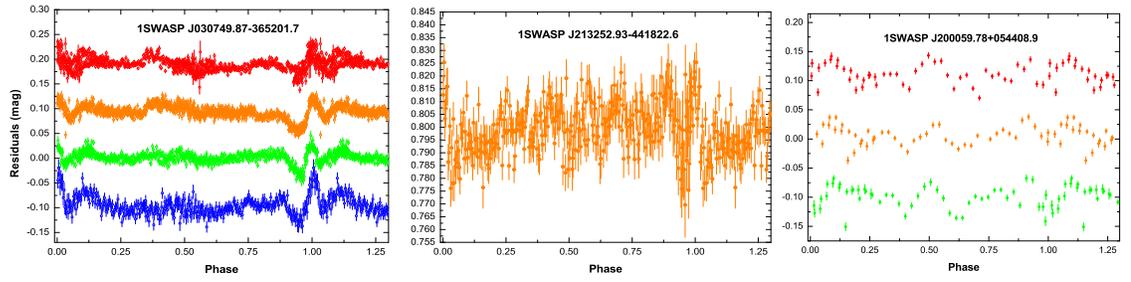}
\caption{Fitting residuals of light curves for the USPCBs. The colors denote the same meaning as the Fig~\ref{fig:light curves_figure}.}
    \label{fig:residuals_figure}
\end{center}
\end{figure}

\begin{figure}
\begin{center}
\includegraphics[angle=0,scale=.85]{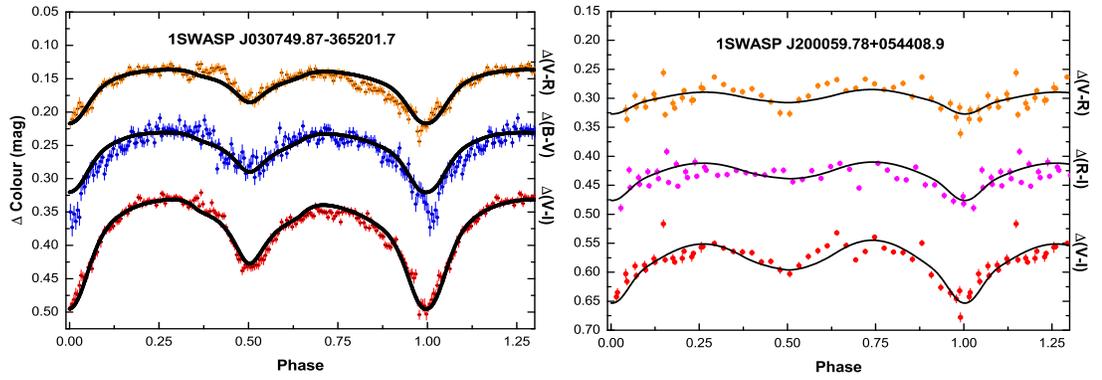}
\caption{Observed differential colors for the three USPCBs. Blue denotes $\Delta$(B-V); orange denotes $\Delta$(V-R); red denotes $\Delta$(V-I); magenta denotes $\Delta$(R-I). The solid lines are yielded by the W-D program.}
    \label{fig:colour index_figure}
\end{center}
\end{figure}

\begin{figure}
\begin{center}
\includegraphics[angle=0,scale=2]{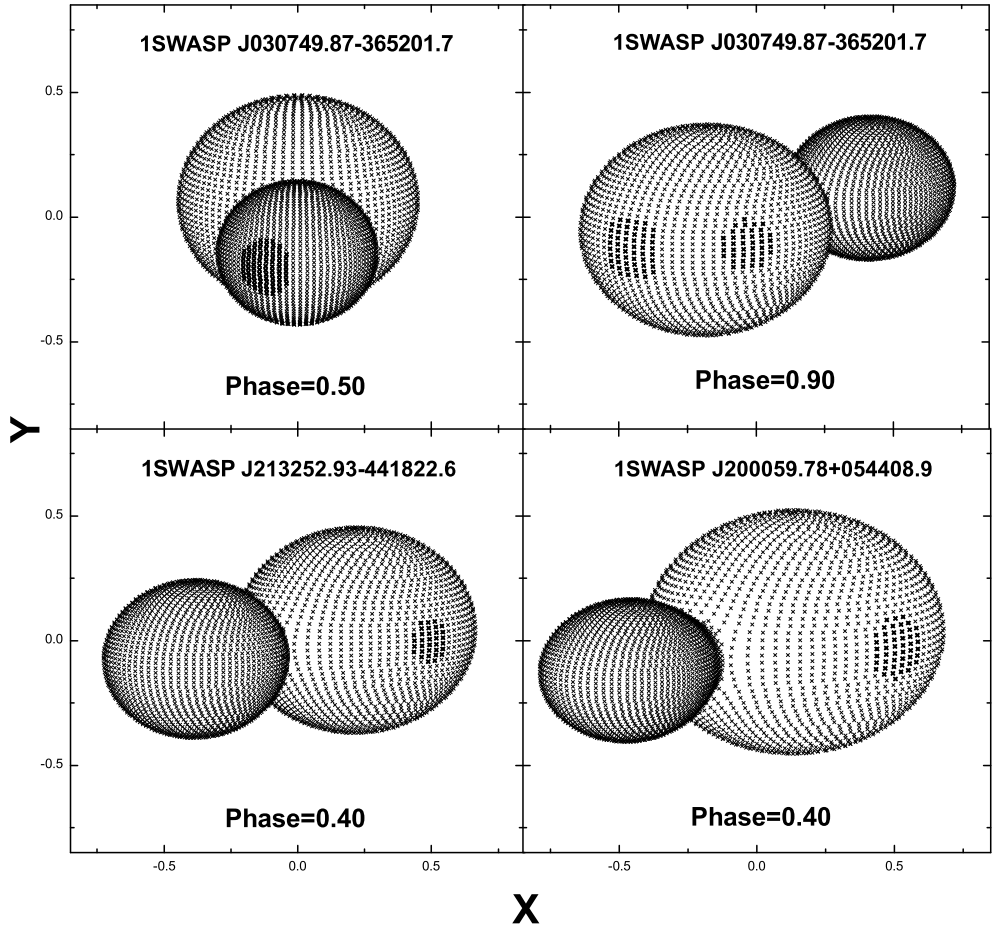}
\caption{The geometrical structure of the three USPCBs with their present cool spots.}
    \label{fig:cool spots_figure}
\end{center}
\end{figure}

Finally, the light curve solutions are listed in Table~\ref{tab:solution_table} and the elements of the assumptive cool spots are listed in Table~\ref{tab:cool spots_table}. The corresponding fittings of each light curve are shown in Fig~\ref{fig:light curves_figure}, and their residuals are shown in Fig~\ref{fig:residuals_figure}. The differential color diagrams are shown in Fig~\ref{fig:colour index_figure}. In this figure, it is clearly seen that the colors of the systems vary with the phases, like their light curves. Particularly, there is a swell in the color curves of J030749 around the phase 0.40. This swell means a higher temperature. The amplitude of the color curves suggests an uncertainty of 700\,K for the temperature estimation, but it does not change any uncertainty of the temperature ratio ($T_1/T_2$). At last, the geometric constructions of the three USPCBs and the present cool spots are displayed in Fig~\ref{fig:cool spots_figure}.

Although our model with cool spots is probably reasonable, and in most situations the differential of results with and without cool spots is usually less than 10\% \citep[e.g.][]{Qianetal2011, Qianetal2013a}, it should be noted that any light curve features whatsoever can be reproduced by the inclusion of sufficient numbers of carefully-tailored spots, and the precise nature of spots can only be established convincingly through a separate method such as Doppler imaging. Hence, for a comparison, we show the unspotted solutions in Table~\ref{tab:unspotted solution_table}, and display the corresponding simulated light curves by dashed lines in Fig~\ref{fig:light curves_figure}.

\section{Discussion of the solutions}
In this section, we will mainly discuss about the reliability of the photometric adjustments and the evolutionary states revealed by these solutions. The resulting inclinations for the three systems are greater than 75 degrees, strongly indicating that their photometric mass ratios are similar to their spectroscopic mass ratios \citep{Maceronivan1996}. Hence, the photometric mass ratios of our solutions should be acceptable. The solutions also show that these three systems are in contact phase, with moderate mass ratios. However, the shortest period system J200059, has a deep contact factor of 58.4\,\%.

Deep USPCBs are very few. This fact is supported both by observational evidence and by theory. The 28 well-studied USPCBs (P $<$ 0.24 day) are collected in Table~\ref{tab:28 well-studied USPCBs_table}. In this table, the mass ratio $q$ is uniformly calculated by $M_s/M_p$, where $M_s$ is the mass of the secondary component (less massive), while $M_p$ is the mass of the primary component (less massive). Only 4 (maybe 5) of these 28 USPCBs have deep fill-out factors ($f>50\,\%$), while other 17 systems are shallow contact ($f<20\,\%$). Unlike the F and G type contact binaries, none of their mass ratios are less than 0.3. Generally, shallow contact factor implies an early phase of contact. According to the thermal relaxation oscillation (TRO) theory \citep{Lucy1976, Flannery1976, Webbink1977}, the material flow starts when the primary component (donor) fully fills its Roche lobe, making the orbit shrink, and then, the secondary component (accretor) fully fills its Roche lobe too. The accretor will expand continuously while it accepts mass, making the orbit wider until the contact configuration is broken. Subsequently, the orbit will shrink again with the mass transfer from the primary to the secondary component. It forms a cycle of contact-semidetached-contact states. In these cycles, the temperatures of the two components will be similar because of the thermal exchange with the mass transfer. The fill-out factor cannot become high when the system undergoes the TRO cycle because of the existence of the contact broken phase. Therefore, TRO explains why it is such a high fraction of shallow USPCBs (17 of 28 systems). For this reason, the zero contact binary within thermal equilibrium J030749, should be in the turn-off phase of the TRO cycle. On the other hand, however, the TRO theory cannot explain the presence of the deep USPCBs with high or moderate mass ratios (e.g. J200059).

If the USPCBs reach a deep fill-out factor, a rapid orbit shrinking mechanism is required. Angular momentum loss (AML) caused by the magnetic stellar winds \citep[e.g.][]{Stepien2006, Stepien2011} may be the mechanism. He pointed out that if there is a third body or the system is located in a dense field, it could lose a lot of AM. An excessive AML can also occur in the pre-main-sequence phase.
In that time, he mentioned that these possibilities, especially the last one, are quite rare according to the observations \citep{Stepien2006}. However, some studies showed that the presence of third bodies is not very rare \citep[e.g.][]{PribullaRucinski2006, DAngeloetal2006, Rucinskietal2007, Qianetal2013b, Qianetal2015b}. Maybe, the few deep USPCBs in Table~\ref{tab:28 well-studied USPCBs_table}, are formed by the rare AML way, e.g. interaction with third bodies. 1SWASP\,J234401.81$-$212229.1, which may be a very deep contact binary system, was found a putative third body by \citet{Lohretal2013b}. The possibility of a third body in J200059 could not be excluded due to the low quality of the O-C diagram, although \citet{Lohretal2013a} did not detect any evidence for period change in J200059. In fact, another system, 1SWASP\,J093010.78$+$533859.5 which had been not found period variations by \citet{Lohretal2013b} either, could be a multiple system \citep{Kooetal2014,Lohretal2015}.

\section{Global parameters correction of the ultrashort-period contact binaries}
The physical parameters of close binaries are sufficiently different from single stars or wide binaries because of the strong mutual gravitational force. For example, the surface gravity accelerations of the components in contact binary systems are observably lower than those of the same masses single main-sequence stars. Hence, to obtain these parameters for close binaries independently, is very important to build an improved evolutionary model for them. It is predicted that USPCBs have a better-constrained empirical global parameter relation than that of F, G and early K type contact binaries because most of them are near the zero age main sequence and/or unevolved. This feature is good for estimating the parameters of USPCBs when spectroscopic data is lacking. \citet{DimitrovKjurkchieva2015} had already summarized a relationship between the period and the semi-major axis, based on the 14 well-studied binaries with P $<$ 0.27 d ($a = -1.154 + 14.633 \times P - 10.319 \times P^2$). However, one point deviated from this relation, i.e., GSC\,1387-0475. It was first investigated by \citet{RucinskiPribulla2008} who obtained a spectroscopic mass ratio.
Unfortunately, their photometric light curve was not of a good quality. \citet{Yangetal2010} observed this system again and obtained $BVR$-band light curves. They adopted the spectroscopic mass ratio given by \citet{RucinskiPribulla2008}, and found smaller masses for the components than theirs. We assumed that this disagreement between both results is caused by the deep fill-out factor. Consequently, we realized that the semi-major axis is probably over estimated with the relation of \citet{DimitrovKjurkchieva2015} when the fill-out factor is huge (e.g. $f>50\,\%$).
According to the well known Roche potential,
\begin{equation}
{\psi}=\frac{2}{1+q}\cdot\frac{1}{r_1}+\frac{2q}{1+q}\cdot\frac{1}{r_2}+(x-\frac{q}{1+q})^2+y^2
	\label{eq:Roche potential}
\end{equation}
where ${r_1}^2=x^2+y^2+z^2$, ${r_2}^2=(1-x)^2+y^2+z^2$, $q=M_2/M_1$, $xyz$ are normalized coordinates (with semi-major axis, A), and
\begin{equation}
f=\frac{{\psi}-{\psi}_{\rm{in}}}{{\psi}_{\rm{out}}-{\psi}_{\rm{in}}},
	\label{eq:fill-out factor}
\end{equation}
is the fill-out factor definition, we computed the $r_{\rm{side}}$, $r_{\rm{back}}$ and $r_{\rm{pole}}$ with a certain fill-out factor $f$ and mass ratio $q$. The effective radius is calculated by
$r_{\rm{E}}=\sqrt[3]{r_{\rm{side}}{\cdot}r_{\rm{back}}{\cdot}r_{\rm{pole}}}$. Then, we compared this $r_{\rm{E}}$ to $r_{\rm{L}}$
\begin{equation}
r_{\rm{L}}=\frac{0.49q^{-2/3}}{0.6q^{-2/3}+\ln(1+q^{-1/3})}.
	\label{eq:critical Roche radii}
\end{equation}
which was provided by \citet{Eggleton1983} with an error less than 1\,\%. Finally, we obtained a ratio of these two normalized radii and used it to correct the semi-major axis. The corresponding results are listed in Table~\ref{tab:Semi-major axis corrections_table}. All errors listed in this table are estimated with the error propagation formula. Our calculated masses of GSC\,1387-0475 are consistent with the values of \citet{Yangetal2010}. If we adopt the corrections for J200059, we find that it can be composed of two M type components with mass $0.458\pm0.066\,M_{\odot}$ and $0.199\pm0.033\,M_{\odot}$, respectively.

\section{Conclusions}
J030749 is a zero contact binary, with $q=0.439\pm0.003$ and $f=0.0\pm3.6\,\%$, placed on the turn-off phase of the TRO cycle. J213252 is a shallow contact binary with $q=0.560\pm0.003$ and $f=14.2\pm1.9\,\%$, under the TRO control. J200059 is a deep contact binary, with $q=0.436\pm0.008$ and $f=58.4\pm1.8\,\%$. It should be formed by a rapid AML mechanism. The three targets show strong magnetic activities. In summary, J030749 and J200059 are in interesting evolutionary stages. It would be worthwhile to monitor them in the future.

%\bigskip
\acknowledgments{
This work is partly supported by the Yunnan Natural Science Foundation (2016FB004), the young academic and technology leaders project of Yunnan Province (No. 2015HB098), National Natural Science Foundation of China (Nos. 11773066, 11403095 and 11325315), the Key Research Programme of the Chinese Academy of Sciences (grant No. KGZD-EW-603), and Strategic Priority Research Programme ``The Emergence of Cosmological Structures'' of the Chinese Academy of Sciences (No.XDB09010202). The new observations were made with the telescopes at Complejo Astron\'{o}mico El Leoncito Observatory (CASLEO), San Juan, Argentina, and the Kottamia Observatory, National Research Institute of Astronomy and Geophysics, Helwan, Cairo, Egypt.

We are also grateful to the anonymous referee who has offered very useful suggestions to improve the paper greatly.
}

%\begin{references}
%\end{references}

\begin{table}
	\centering
\begin{footnotesize}
\caption{Summary of the observations.}
	\label{tab:summary observation_table}
\begin{tabular}{ccccccc}
\hline
        Target	               &  Obs Date	& Texp (s)	    & Filter	    &N Img	&Telescope	    &Seeing ($''$)\\
\hline
1SWASP\,J030749.87$-$365201.7  & 2014-11-24	& 30	        &   I	        & 471	&2.15\,m JST	&3.0-7.0   \\
	                           & 2014-11-25	& 30	        &   I	        & 440	&2.15\,m JST	&3.5-7.5   \\
	                           & 2014-12-03	& 30	        &   V	        & 669	&2.15\,m JST	&3.5-7.5   \\
	                           & 2014-12-04	& 60	        &   B	        & 325	&2.15\,m JST	&3.5-7.0   \\
1SWASP\,J213252.93$-$441822.6  & 2015-09-11	& 60	        &   R	        & 308	&2.15\,m JST	&3.4-7.4   \\
1SWASP\,J200059.78$+$054408.9  & 2016-08-01	& 200	        &   V	        & 49	&1.88\,m KRT	&3.0-6.5   \\
	                           & 2016-08-01	& 135	        &   R	        & 46	&1.88\,m KRT	&3.0-6.5   \\
	                           & 2016-08-01	& 170	        &   I	        & 46	&1.88\,m KRT	&3.0-6.5   \\
\hline
\end{tabular}
\end{footnotesize}
\end{table}

\begin{table}
	\centering
\begin{tiny}
\caption{The JHK magnitudes for the targets, comparisons and check stars. Estimated temperatures of the primary components for the three USPCBs based on the JHK colors.}
	\label{tab:target comparison_table}
\begin{tabular}{lllllll}
\hline
Name				&  J (mag) &  H (mag) & K (mag) &  J-K & Type        &$T_p$ (K)\\
\hline
1SWASP\,J030749.87$-$365201.7	&13.552    &13.007    &12.887 	&0.665  &Target	     &4750  \\
2MASS\,J03075380$-$3653319	    &12.840    &12.407    &12.359 	&0.481  &Comparison  &      \\
2MASS\,J03075601$-$3652174	    &15.165    &14.910    &14.873 	&0.292  &Check	     &      \\
1SWASP\,J213252.93$-$441822.6	&14.299    &13.667    &13.622 	&0.677  &Target	     &4700  \\
2MASS\,J21325995$-$4418032	    &12.782    &12.317    &12.228 	&0.554  &Comparison  &      \\
2MASS\,J21330002$-$4418197	    &13.652    &13.263    &13.178 	&0.474  &Check	     &      \\
1SWASP\,J200059.78$+$054408.9	&13.412    &12.875    &12.764 	&0.648  &Target	     &4800  \\
2MASS\,J20011488$+$0543026	    &13.519    &13.162    &13.048 	&0.471  &Comparison  &      \\
2MASS\,J20011491$+$0542524	    &13.420    &13.204    &13.225 	&0.195  &Check	     &      \\
\hline
\end{tabular}
\end{tiny}
\end{table}

\begin{table}
	\centering
\begin{tiny}
\caption{Spotted photometric solutions for the three USPCBs.}
	\label{tab:solution_table}
\begin{tabular}{lclclcl}
\hline
                        &  1SWASP\,J030749.87$-$365201.7   &	 &  1SWASP\,J213252.93$-$441822.6    &       &  1SWASP\,J200059.78$+$054408.9          \\
                                                                                                                                                       \\
Parameters              &  Photometric elements  &  errors       &  Photometric elements  &  errors          &  Photometric elements  &  errors        \\
\hline                                                                                                                                                 \\
$g_1=g_2$               &     0.32               & assumed       &     0.32               & assumed          &     0.32               & assumed        \\
$A_1=A_2$               &     0.50               & assumed       &     0.50               & assumed          &     0.50               & assumed        \\
$x_{1bolo}=x_{2bolo}$   &     0.315              & assumed       &     0.315              & assumed          &     0.311              & assumed        \\
$y_{1bolo}=y_{2bolo}$   &     0.371              & assumed       &     0.370              & assumed          &     0.377              & assumed        \\
$x_{1V}=x_{2B}$         &     1.036              & assumed       &     --                 & --               &     --                 & --             \\
$y_{1V}=y_{2B}$         &   $-0.216$             & assumed       &     --                 & --               &     --                 & --             \\
$x_{1V}=x_{2V}$         &     0.682              & assumed       &     --                 & --               &     0.658              & assumed        \\
$y_{1V}=y_{2V}$         &     0.135              & assumed       &     --                 & --               &     0.162              & assumed        \\
$x_{1R}=x_{2R}$         &     0.423              & assumed       &     0.437              & assumed          &     0.407              & assumed        \\
$y_{1R}=y_{2R}$         &     0.343              & assumed       &     0.330              & assumed          &     0.360              & assumed        \\
$x_{1I}=x_{2I}$         &     0.258              & assumed       &     --                 & --               &     0.247              & assumed        \\
$y_{1I}=y_{2I}$         &     0.423              & assumed       &     --                 & --               &     0.433              & assumed        \\
Phase shift             &     --                 & --            &     --                 & --               &     0.0032             & $\pm0.0006$    \\
$T_1$ (K)               &     4750               & $\pm700$      &     4700               & $\pm400$         &     4800               & $\pm300$       \\
$T_2$ (K)               &     4697               & $\pm703$  	 &     4671               & $\pm405$         &     4528               & $\pm316$       \\
$q=M_2/M_1$             &     2.280              & $\pm0.010$	 &     1.785              & $\pm0.007$       &     2.296              & $\pm0.040$     \\
$\Omega_{in}$           &     5.6432             & --     		 &     4.9443             & --               &     5.6660             & --             \\
$\Omega_{out}$          &     5.0383             & --     		 &     4.3545             & --               &     5.0606             & --             \\
$\Omega_1=\Omega_2$     &     5.6431             & $\pm0.0220$   &     4.8609             & $\pm0.0110$	     &     5.3123             & $\pm0.0112$    \\ 	
$i (^{\circ})$          &     $78.2$             & $\pm0.1$ 	 &     81.9               & $\pm0.1$         &     $75.3$             & $\pm0.5$       \\
$L_1/(L_1+L_2)(B)$      &     0.3415             & $\pm0.0015$ 	 &     --                 & --               &     --                 & --             \\
$L_1/(L_1+L_2)(V)$      &     0.3231             & $\pm0.0012$ 	 &     --                 & --               &     0.4196             & $\pm0.0067$    \\
$L_1/(L_1+L_2)(R)$      &     0.3035             & $\pm0.0010$ 	 &     0.3789             & $\pm0.0013$      &     0.3975             & $\pm0.0054$    \\
$L_1/(L_1+L_2)(I)$      &     0.2801           	 & $\pm0.0009$	 &     --                 & --   		     &     0.3850             & $\pm0.0048$    \\
$r_1(pole)$             &     0.2951             & $\pm0.0014$	 &     0.3165             & $\pm0.0012$ 	 &     0.3200             & $\pm0.0064$    \\
$r_1(side)$             &     0.3079             & $\pm0.0017$	 &     0.3319             & $\pm0.0015$      &     0.3392             & $\pm0.0082$    \\
$r_2(back)$             &     0.3402             & $\pm0.0027$	 &     0.3692             & $\pm0.0025$      &     0.4027             & $\pm0.0195$    \\
$r_2(pole)$             &     0.4185             & $\pm0.0013$	 &     0.4124             & $\pm0.0011$      &     0.4547             & $\pm0.0049$    \\
$r_2(side)$             &     0.4449             & $\pm0.0016$   &     0.4386             & $\pm0.0014$      &     0.4924             & $\pm0.0069$    \\
$r_2(back)$             &     0.4724             & $\pm0.0021$   &     0.4704             & $\pm0.0020$      &     0.5333             & $\pm0.0102$    \\
$f$ (\%)                &     0.0                & $\pm3.6$		 &     14.2               & $\pm1.9$	     &     58.4               & $\pm1.8$	   \\
\hline
\end{tabular}
\end{tiny}
\end{table}

\begin{table}
	\centering
\begin{tiny}
\caption{Cool spot elements based on the spotted light curve solutions.}
	\label{tab:cool spots_table}
\begin{tabular}{lccccc}
\hline
                               &          & $\theta$ ($^\circ$) &  $\psi$ ($^\circ$) &  $\Omega($sr$)$ &  $T_s/T_*$\\
\hline
1SWASP\,J030749.87$-$365201.7  & star 1   &  89.95              &  157.08            &    0.35845      &    0.700  \\
                               & star 2   &  86.51              &  235.49            &    0.22845      &    0.850  \\
                               & star 2   &  89.95              &  180.00            &    0.27845      &    0.700  \\
1SWASP\,J213252.93$-$441822.6  & star 2   &  89.67              &   77.01            &    0.20800      &    0.700  \\
1SWASP\,J200059.78$+$054408.9  & star 2   &  89.70              &  134.30            &    0.20845      &    0.700  \\
\hline
\end{tabular}
\end{tiny}
\end{table}

\begin{table}
	\centering
\begin{tiny}
\caption{Unspotted photometric solutions for the three USPCBs.}
	\label{tab:unspotted solution_table}
\begin{tabular}{lclclcl}
\hline
                        &  1SWASP\,J030749.87$-$365201.7   &	 &  1SWASP\,J213252.93$-$441822.6    &       &  1SWASP\,J200059.78$+$054408.9          \\
                                                                                                                                                       \\
Parameters              &  Photometric elements  &  errors       &  Photometric elements  &  errors          &  Photometric elements  &  errors        \\
\hline                                                                                                                                                 \\
$g_1=g_2$               &     0.32               & assumed       &     0.32               & assumed          &     0.32               & assumed        \\
$A_1=A_2$               &     0.50               & assumed       &     0.50               & assumed          &     0.50               & assumed        \\
$x_{1bolo}=x_{2bolo}$   &     0.315              & assumed       &     0.315              & assumed          &     0.311              & assumed        \\
$y_{1bolo}=y_{2bolo}$   &     0.371              & assumed       &     0.370              & assumed          &     0.377              & assumed        \\
$x_{1V}=x_{2B}$         &     1.036              & assumed       &     --                 & --               &     --                 & --             \\
$y_{1V}=y_{2B}$         &   $-0.216$             & assumed       &     --                 & --               &     --                 & --             \\
$x_{1V}=x_{2V}$         &     0.682              & assumed       &     --                 & --               &     0.658              & assumed        \\
$y_{1V}=y_{2V}$         &     0.135              & assumed       &     --                 & --               &     0.162              & assumed        \\
$x_{1R}=x_{2R}$         &     0.423              & assumed       &     0.437              & assumed          &     0.407              & assumed        \\
$y_{1R}=y_{2R}$         &     0.343              & assumed       &     0.330              & assumed          &     0.360              & assumed        \\
$x_{1I}=x_{2I}$         &     0.258              & assumed       &     --                 & --               &     0.247              & assumed        \\
$y_{1I}=y_{2I}$         &     0.423              & assumed       &     --                 & --               &     0.433              & assumed        \\
Phase shift             &     --                 & --            &     --                 & --               &     0.0032             & $\pm0.0006$    \\
$T_1$ (K)               &     4750               & $\pm700$      &     4700               & $\pm400$         &     4800               & $\pm300$       \\
$T_2$ (K)               &     4630               & $\pm704$  	   &     4706               & $\pm406$         &     4532               & $\pm323$       \\
$q=M_2/M_1$             &     2.113              & $\pm0.010$	   &     1.783              & $\pm0.015$       &     2.358              & $\pm0.055$     \\
$\Omega_{in}$           &     5.4107             & --     		   &     4.9419             & --               &     5.7514             & --             \\
$\Omega_{out}$          &     4.8102             & --     		   &     4.3521             & --               &     5.1449             & --             \\
$\Omega_1=\Omega_2$     &     5.4050             & $\pm0.0220$   &     4.8653             & $\pm0.0033$	     &     5.4039              & $\pm0.0187$    \\ 	
$i (^{\circ})$          &     $78.2$             & $\pm0.1$ 	   &     81.9               & $\pm0.2$         &     $74.7$             & $\pm0.6$       \\
$L_1/(L_1+L_2)(B)$      &     0.3803             & $\pm0.0016$ 	 &     --                 & --               &     --                 & --             \\
$L_1/(L_1+L_2)(V)$      &     0.3709             & $\pm0.0012$ 	 &     --                 & --               &     0.4196             & $\pm0.0067$    \\
$L_1/(L_1+L_2)(R)$      &     0.3618             & $\pm0.0009$ 	 &     0.3940             & $\pm0.0020$      &     0.3975             & $\pm0.0054$    \\
$L_1/(L_1+L_2)(I)$      &     0.3566             & $\pm0.0007$	 &     --                 & --   		         &     0.3850             & $\pm0.0048$    \\
$r_1(pole)$             &     0.2960             & $\pm0.0014$	 &     0.3160             & $\pm0.0027$ 	   &     0.3169             & $\pm0.0089 $   \\
$r_1(side)$             &     0.3089             & $\pm0.0017$	 &     0.3312             & $\pm0.0033$      &     0.3357             & $\pm0.0114 $   \\
$r_2(back)$             &     0.3417             & $\pm0.0027$	 &     0.3681             & $\pm0.0055$      &     0.3981             & $\pm0.0269 $   \\
$r_2(pole)$             &     0.4193             & $\pm0.0013$	 &     0.4116             & $\pm0.0024$      &     0.4559             & $\pm0.0070 $   \\
$r_2(side)$             &     0.4459             & $\pm0.0016$   &     0.4377             & $\pm0.0031$      &     0.4937             & $\pm0.0098 $   \\
$r_2(back)$             &     0.4738             & $\pm0.0022$   &     0.4692             & $\pm0.0043$      &     0.5337             & $\pm0.0143 $   \\
$f$ (\%)                &     1.0                & $\pm3.7$		   &     13.0               & $\pm4.2$	       &     57.3               & $\pm1.8$	   \\
\hline
\end{tabular}
\end{tiny}
\end{table}

\clearpage
\begin{table}
\begin{tiny}
\begin{center}
\caption{28 well-studied USPCBs.}
	\label{tab:28 well-studied USPCBs_table}
\begin{tabular}{lccccccccl}
\hline
Name                            &   Period (day)&$q=M_s/M_p$& $f (\%)$& $i (^{\circ})$& $M_p^*$ & $M_s^*$ & $R_p^*$ & $R_s^*$ & reference                       \\
\hline
1SWASP\,J030749.87$-$365201.7   &   0.2266712   &   0.439  &  0.0   &  78.2 & 0.789 	&0.346 	&0.737 	&0.507 & This paper                               \\
RW Com                          &   0.2373464   &   0.471  &  6.1   &  74.9 & 0.849 	&0.400 	&0.774 	&0.549 & \citet{Djurasevicetal2011}                   \\
NSVS\,2700153                   &   0.228456    &   0.775  &  7.1   &  47.8 & 0.650 	&0.504 	&0.662 	&0.589 & \citet{DimitrovKjurkchieva2015}            \\
1SWASP\,J055416.98$+$442534.0   &   0.21825     &   0.792  &  8.6   &  70.1 & 0.582 	&0.461 	&0.618 	&0.556 & \citet{DimitrovKjurkchieva2015}            \\
1SWASP\,J200503.05$-$343726.5   &   0.2288836   &   0.934  &  9.0   &  73.8 & 0.599 	&0.560 	&0.637 	&0.617 & \citet{Zhangetal2017}                        \\
1SWASP\,J160156.04$+$202821.6   &   0.22653     &   0.670  &  10.0  &  79.5 & 0.679 	&0.455 	&0.675 	&0.563 & \citet{Lohretal2014, Essametal2014}   \\
1SWASP\,J022727.03$+$115641.7   &   0.21095     &   0.463  &  10.4  &       & 0.659 	&0.305 	&0.658 	&0.464 & \citet{Liuetal2015a}                       \\
1SWASP\,J150822.80$-$054236.9   &   0.23006     &   0.514  &  12.0  &  90.0 & 0.774 	&0.398 	&0.729 	&0.538 & \citet{Lohretal2014}                       \\
1SWASP\,J074658.62$+$224448.5   &   0.22085     &   0.395  &  12.6  &  78.6 & 0.769 	&0.304 	&0.726 	&0.476 & \citet{Jiangetal2015a}                      \\
1SWASP\,J080150.03$+$471433.8   &   0.217531    &   0.432  &  13.6  &  83.8 & 0.723 	&0.313 	&0.698 	&0.476 & \citet{DimitrovKjurkchieva2015}            \\
1SWASP\,J213252.93$-$441822.6   &   0.22123459  &   0.560  &  14.2  &  81.9 & 0.690 	&0.386 	&0.679 	&0.521 & This paper                               \\
1SWASP\,J015100.23$-$100524.2   &   0.2145001   &   0.320  &  14.6  &  79.4 & 0.760 	&0.243 	&0.724 	&0.432 & \citet{Qianetal2015b}                         \\
V1104 Her                       &   0.22788     &   0.623  &  15.0  &       & 0.707 	&0.441 	&0.692 	&0.557 & \citet{Liuetal2015b}                       \\
CC Com                          &   0.22068516  &   0.526  &  17.0  &  89.8 & 0.701 	&0.369 	&0.685 	&0.511 & \citet{Koseetal2011}                         \\
1SWASP\,J093010.78$+$533859.5B  &   0.22771377  &   0.397  &  17.0  &  86.0 & 0.821 	&0.326 	&0.757 	&0.497 & \citet{Lohretal2015}           \\
SDSS\,J012119.10$¨C$001949.9    &   0.2052      &   0.500  &  18.9  &  83.9&  0.600 	&0.300 	&0.622 	&0.454  & \citet{Jiangetal2015b}                     \\
NSVS\,7179685                   &   0.20974     &   0.451  &  19.3  &  85.5 & 0.655 	&0.295 	&0.656 	&0.457 & \citet{DimitrovKjurkchieva2015}            \\
NSVS\,8626028                   &   0.217407    &   0.805  &  20.7  &  65.9 & 0.573 	&0.461 	&0.612 	&0.555 & \citet{DimitrovKjurkchieva2015}            \\
2MASS\,J00164102$-$0009251      &   0.198561    &   0.630  &  22.0  &  53.3 & 0.507 	&0.319 	&0.564 	&0.457 & \citet{Davenportetal2013}                    \\
1SWASP\,J024148.62$+$372848.3   &   0.21975076  &   0.813  &  23.3  &  68.7 & 0.585 	&0.475 	&0.621 	&0.565 & \citet{Jiangetal2015c}                      \\
1SWASP\,J133105.91$+$121538.0   &   0.21801     &   0.828  &  25.2  &  77.6 & 0.570 	&0.472 	&0.611 	&0.561 & \citet{Elkhateebetal2014a}                 \\
V523 Cas                        &   0.233693    &   0.516  &  29.0  &  85.4 & 0.798 	&0.412 	&0.744 	&0.550 & \citet{Samecetal2004}                        \\
1SWASP\,J210318.76$+$021002.2   &   0.22859     &   0.877  &  34.2  &  81.9 & 0.616 	&0.540 	&0.645 	&0.607 & \citet{Elkhateebetal2014b}                 \\
1SWASP\,J200059.78$+$054408.9   &   0.20569054  &   0.435  &  58.4  &  75.3 & 0.631 	&0.274 	&0.642 	&0.439 & This paper                               \\
NSVS\,925605                    &   0.217629    &   0.678  &  70.2  &  57.2 & 0.618 	&0.419 	&0.637 	&0.533 & \citet{DimitrovKjurkchieva2015}            \\
GSC\,1387$-$0475                &   0.21781128  &   0.474  &  76.3  &  49.9 & 0.705 	&0.334 	&0.687 	&0.489 & \citet{Yangetal2010}                         \\
1SWASP\,J075102.16$+$342405.3   &   0.20917224  &   0.740  &  96.0  &  76.0 & 0.543 	&0.401 	&0.590 	&0.514 & \citet{Jiangetal2015d}                      \\
1SWASP\,J234401.81$-$212229.1   &   0.21367     &   0.422  &  deep? &  79.4 & 0.699 	&0.295 	&0.683 	&0.461 & \citet{Lohretal2013b, Koen2014}           \\
\hline
\end{tabular}
\end{center}
\noindent {$^*$ Computed by the relation of \citet{DimitrovKjurkchieva2015} and by equation~\ref{eq:critical Roche radii}, in solar units. The footnote P denotes the primary component, while S denotes the secondary component.}
\end{tiny}
\end{table}

\begin{table}
	\centering
\begin{tiny}
\caption{Semi-major axis corrections for the deep USPCBs.}
	\label{tab:Semi-major axis corrections_table}
\begin{tabular}{lcccc}
\hline
Name                                                         & 1SWASP\,J200059.78$+$054408.9  & NSVS\,925605     & GSC\,1387$-$0475       & 1SWASP\,J075102.16$+$342405.3     \\
Period (day)                                                 & 0.20569054 	                  & 0.21762900       & 0.21781128 	          & 0.20917224                        \\
$q = M_s/M_p$                                                & $0.436\pm0.008$ 	              & $0.678\pm0.003$  & $0.474\pm0.008$        & $0.740\pm0.040$                   \\
$f(\%)$                                                      & $58.4\pm1.8$ 	              & $70.2\pm2.6$ 	 & $76.3\pm2.9$ 	      & $95.0\pm4.0$                      \\
Refrence                                                     &This paper                      &\citet{DimitrovKjurkchieva2015} & \citet{Yangetal2010} & \citet{Jiangetal2015d}\\
\hline
a $(R_{\odot})$                                              & $1.419\pm0.050$ 	              & $1.542\pm0.050$  & $1.544\pm0.050$ 	      & $1.455\pm0.050$                   \\
$r_{\rm{L1}}$                                                & $0.452\pm0.005$  	          & $0.413\pm0.004$  & $0.445\pm0.004$  	  & $0.405\pm0.004$                   \\
$r_{\rm{L2}}$                                                & $0.310\pm0.003$  	          & $0.346\pm0.003$  & $0.316\pm0.003$  	  & $0.353\pm0.004$                   \\
$r_{\rm{E1}}$                                                & $0.492\pm0.011$  	          & $0.475\pm0.012$  & $0.502\pm0.012$  	  & $0.505\pm0.012$                   \\
$r_{\rm{E2}}$                                                & $0.352\pm0.010$  	          & $0.412\pm0.010$  & $0.380\pm0.011$  	  & $0.467\pm0.011$                   \\
$cor=(r_{\rm{L1}}/r_{\rm{E1}}+r_{\rm{L2}}/r_{\rm{E2}})/2$    & $0.899\pm0.013$   	          & $0.854\pm0.013$  & $0.860\pm0.015$  	  & $0.780\pm0.010$                   \\
$a'=a \cdot cor (R_{\odot})$                                 & $1.276\pm0.064$ 	              & $1.317\pm0.063$  & $1.327\pm0.066$ 	      & $1.135\pm0.054$                   \\
$M_p (M_{\odot})$                                            & $0.458\pm0.066$  	          & $0.385\pm0.055$  & $0.448\pm0.064$ 	      & $0.257\pm0.031$                   \\
$M_s (M_{\odot})$                                            & $0.199\pm0.033$  	          & $0.261\pm0.038$  & $0.212\pm0.034$ 	      & $0.190\pm0.033$                   \\
$R_p=r_{\rm{E1}} \cdot a' (R_{\odot})$                       & $0.628 \pm0.046$  	          & $0.626\pm0.046$  & $0.666\pm0.049$  	  & $0.572\pm0.041$                   \\
$R_s=r_{\rm{E2}} \cdot a' (R_{\odot})$                       & $0.449 \pm0.035$  	          & $0.543\pm0.039$  & $0.504\pm0.040$  	  & $0.530\pm0.038$                   \\
\hline
\end{tabular}
\end{tiny}
\end{table}

\clearpage

\end{document}